\documentclass{caosp306}
\usepackage{graphicx}

\usepackage{natbib}
\usepackage{xcolor}
\usepackage{amsmath}
\usepackage{amssymb}
\def\pasa{PASA}%
\def\zap{ZAp}%
\def\eprint[#1]{#1}
\articleNo{0}
\pubyear{2020}
\volume{50}
\volnumber{0}
\firstpage{472}
\received{December 2, 2019}
\accepted{January 26, 2020}

\def\BibTeX{{\rm B\kern-.05em{\sc i\kern-.025em b}\kern-.08em
             T\kern-.1667em\lower.7ex\hbox{E}\kern-.125emX}}
         
\begin{document}

%%%%%%%%%%%%%%%%%%%%%%%%%%%%%%%%%%%%%%%%%%%%%%%%%%%%%%%%%%%%%%%%%%%%%%%%%%%%%
%              R U N N I N G   P A G E   H E A D I N G S                     
% Odd page headings (except for the title page) are produced automatically
% and contain the title. If, and only if, the title of your article is too
% long the running head is omitted in the printout; you can make your own
% running title by using the \htitle command, putting the shortened title
% between the curly brackets. \htitle should also be used when the
% subtitle is present: \htitle offers you a way how to include it into the
% headings. If you wish to see how it works simply remove the % sign from
% the beginning of that line.
%
% Unlike the \htitle command, the \hauthor command is compulsory. It is
% used to produce even page headings and contains the names of the authors
% of an article.  All authors must be listed here, if possible. When
% authors' list is too long, you can abbreviate it by using "{\it et
% al.}". Authors' names are given in the form: initial(s) of the author's
% first name and surname. Authors are separated by a "," (comma) sign and
% the last one by "and".
%%%%%%%%%%%%%%%%%%%%%%%%%%%%%%%%%%%%%%%%%%%%%%%%%%%%%%%%%%%%%%%%%%%%%%%%%%%%%
\htitle{The explosive life of massive binaries}
\hauthor{M.~Renzo \& E.~Zapartas}

%%%%%%%%%%%%%%%%%%%%%%%%%%%%%%%%%%%%%%%%%%%%%%%%%%%%%%%%%%%%%%%%%%%%%%%%%%%%%
%                       T I T L E                                            
% Capital letters in the title are only used at the beginning of the
% names. Don`t end the title by a "." (dot)
%%%%%%%%%%%%%%%%%%%%%%%%%%%%%%%%%%%%%%%%%%%%%%%%%%%%%%%%%%%%%%%%%%%%%%%%%%%%%
\title{The explosive life of massive binaries}

%%%%%%%%%%%%%%%%%%%%%%%%%%%%%%%%%%%%%%%%%%%%%%%%%%%%%%%%%%%%%%%%%%%%%%%%%%%%%
%                       S U B T I T L E                                      
% You can use the subtitle, with the command \subtitle similar to the
% \title command.
%%%%%%%%%%%%%%%%%%%%%%%%%%%%%%%%%%%%%%%%%%%%%%%%%%%%%%%%%%%%%%%%%%%%%%%%%%%%%

%%%%%%%%%%%%%%%%%%%%%%%%%%%%%%%%%%%%%%%%%%%%%%%%%%%%%%%%%%%%%%%%%%%%%%%%%%%%%
%                   A U T H O R  N A M E S                                   
% Authors' names are separated by the \and command and their institutes
% are assigned by the \inst{n} command.
%
% When the name contains "Slovak" letters L,d,t,l with a caron, use an
% a new \softl, etc. command (examples given in the last table of
% this document) to produce typographically correct accented characters.
%%%%%%%%%%%%%%%%%%%%%%%%%%%%%%%%%%%%%%%%%%%%%%%%%%%%%%%%%%%%%%%%%%%%%%%%%%%%%
\author{M.~Renzo\inst{1,2} \and E.~Zapartas\inst{3}}

%%%%%%%%%%%%%%%%%%%%%%%%%%%%%%%%%%%%%%%%%%%%%%%%%%%%%%%%%%%%%%%%%%%%%%%%%%%%%
%           I N S T I T U T E S'  A D D R E S S E S                          
% The affiliation of authors is generated by the \institute command, the
% \and command being again used to separate individual addresses.
% The following commands may be used for the following three institutes:   
%               \lomnica        for      AsU SAV, Tatranska Lomnica          
%               \blava          for      AsU SAV, Bratislava                 
%               \ondrejov       for      AsU CAV, Ondrejov                   
%
% The given postal address must be complete in order to facilitate our
% editorial work. Moreover, you can add your e-mail address, using the
% \email command.
%%%%%%%%%%%%%%%%%%%%%%%%%%%%%%%%%%%%%%%%%%%%%%%%%%%%%%%%%%%%%%%%%%%%%%%%%%%%%
\institute{Center for Computational Astrophysics, Flatiron Institute, New York, NY 10010, USA \and
  Anton Pannekoek Institute for Astronomy and Grappa, University of
  Amsterdam, NL-1090 GE Amsterdam, The Netherlands \and  
  Geneva Observatory, University of Geneva, CH-1290 Sauverny, Switzerland}
%%%%%%%%%%%%%%%%%%%%%%%%%%%%%%%%%%%%%%%%%%%%%%%%%%%%%%%%%%%%%%%%%%%%%%%%%%%%%
%                        D A T E / R E C E I V E D                          
% Date inserted here will be the date when your paper was received The
% format is: month (not abbreviated), day, year.
%%%%%%%%%%%%%%%%%%%%%%%%%%%%%%%%%%%%%%%%%%%%%%%%%%%%%%%%%%%%%%%%%%%%%%%%%%%%%
\date{September 10, 2019}

%%%%%%%%%%%%%%%%%%%%%%%%%%%%%%%%%%%%%%%%%%%%%%%%%%%%%%%%%%%%%%%%%%%%%%%%%%%%%
%                        M A K E T I T L E
% The beginning part (title, author(s), etc.) of your article must be
% closed by the \maketitle command.
%%%%%%%%%%%%%%%%%%%%%%%%%%%%%%%%%%%%%%%%%%%%%%%%%%%%%%%%%%%%%%%%%%%%%%%%%%%%%
\maketitle

%%%%%%%%%%%%%%%%%%%%%%%%%%%%%%%%%%%%%%%%%%%%%%%%%%%%%%%%%%%%%%%%%%%%%%%%%%%%%
%                        A B S T R A C T,  K E Y W O R D S                   
% Here it is shown how to write an abstract.  Keywords should be placed
% within the "abstract" environment using the command \keywords and they
% should be selected from the thesaurus from Astron.  Astrophys.
% Abstracts. They must be separated from each other by -- (two dashes).
%%%%%%%%%%%%%%%%%%%%%%%%%%%%%%%%%%%%%%%%%%%%%%%%%%%%%%%%%%%%%%%%%%%%%%%%%%%%%
\begin{abstract}
  Massive stars are born predominantly as members of binary (or higher multiplicity) systems, and the
  presence of a companion can significantly alter their life and final fate. Therefore, any observed
  sample of massive stars or associated transients is likely to be significantly influenced by the
  effects of binarity. Here, we focus on the relationship between massive binary evolution and
  core-collapse supernova events. In the vast majority of the cases, the first core-collapse event
  happening in a binary system unbinds the two stars. Studying the population of companion stars,
  either at the supernova site, or as ``widowed'' stars long after the explosion,
  can be used to constrain the previous orbital evolution of the binary progenitor,
  and explosion physics of their former companion. Specifically, the population of ``widowed'' stars
  might provide statistical constraints on the
  typical amplitude of black hole natal kicks without seeing neither
  the black holes nor the transient possibly associated to their formation. Binarity also has a
  large impact on the predicted population of supernova sub-types, including hydrogen-rich type II
  supernovae, with a significant fraction of hydrogen-rich stars at explosions being either merger
  products or accretors. 
\end{abstract}

\section{Massive stars and binarity}

A variety of observations suggest that the vast majority of massive stars are
born in binary  \citep[e.g.,][]{mason:09, sana:11, almeida:17} or higher multiplicity systems \citep[e.g.,][]{tokovinin:08}, and that up
to $\sim$\,$70\%$ of the O-type stars might exchange mass or merge with a companion
before the end of their evolution \citep[e.g.,][]{sana:12}.

This implies that observational samples
of massive stars and/or transients is likely to contain
binary evolution products (e.g., \citealt{langer:12} for a review and \citealt{demink:14}).

In the era of large surveys, such as \emph{Gaia} \citep[][]{brown:18} or ZTF \citep[][]{bellm:14} and LSST \citep[][]{LSSThandbook} in
the time domain, we have the opportunity to investigate both the
most common \emph{and} the rare and exotic binary evolution paths.

Here, we focus on some aspects of the relationship between binarity and core-collapse supernova (CCSN) explosions.
We refer interested readers for more details to our studies in
\cite{renzo:19walk, zapartas:17,zapartas:17b, zapartas:19}. We describe the population synthesis
approach in Sec.~\ref{sec:pop_synth}. In Sec.~\ref{sec:expl_bin}, we focus on the consequences of the
first CCSN explosion for the binary system, while in Sec.~\ref{sec:bin_expl} we discuss the
implications of binary evolution of the progenitors for the population of CCSN events.

\subsection{Population synthesis}
\label{sec:pop_synth}

The evolution of a star is determined mainly by its initial mass, and secondly by its rotation rate
and metallicity\footnote{specifically, its initial iron content, e.g., \citealt{tramper:16}.} (Z). Nevertheless,
many uncertain or unknown parameters enter in the modeling of internal processes in stars (poorly
known nuclear reaction rates, modeling of mixing processes, wind
mass loss rates, etc.). Exploring the parameter space for single star evolution is a challenging
task still being actively pursued \citep[e.g.,][for recent studies of single massive
  stars]{woosley:17, renzo:17, sukhbold:18, woosley:19, farmer:19}. When considering the
evolution of two stars born together in a binary, i.e., the standard for massive stars, the number
of dimensions of the parameter space increases very rapidly: not only the evolution of a massive
binary system depends on the two initial masses of the stars, but also their initial orbital period
and, possibly, eccentricity. Moreover, the number of free or poorly constrained parameters entering
in the models also increases (e.g., stability and efficiency of mass transfer, angular momentum losses,
treatment of common envelope), reflecting the current insufficient understanding of the physics of
binary interactions.

The complexity of the problem and the vastness of the parameter space to explore require to resort
to population synthesis techniques, i.e., broadly speaking, building synthetic populations by
weighting with initial distributions pre-computed models. This can be done with grids of detailed
binary evolution models \citep[e.g.,][]{justham:14, marchant:16, eldridge:17}, which have
the advantage of solving the differential equations describing the evolution and interaction of the
two stars, but are limited to the values of unknown parameters for which the computation is
numerically feasible.

The common alternative is to rely on pre-computed single star models \citep[e.g.,][]{pols:98,
  hurley:00} paired with analytic algorithms to represent the binary interactions
\citep[e.g.,][]{tout:97,hurley:02}, which allow for the exploration of larger portions of the
parameter space of binary evolution at the cost of a reduced physical accuracy of the models, and
thus limited predictive power of a single population. The results we present here were obtained
with this approach, using the \texttt{binary\_c} code
\citep[][]{izzard:04,izzard:06,izzard:09, izzard:18}. 

The speed of these simulations
(typically $\lesssim 0.1\,\mathrm{sec}$ per binary) allows for re-runs varying the uncertain
parameters. If a particular result is found to survive all the parameter variations possible, the
prediction can be considered robust.
Viceversa, if a result is found to be sensitive to variations of a particular unknown parameter, the
comparison with observed populations has the potential of constraining such parameter. The ranges
reported below were obtained with one-by-one variations of uncertain parameters\footnote{Note however that this approach neglects possible
  physical correlations between the parameters, see also, e.g., \citealt{andrews:18,taylor:18}.} in
the models to assess the robustness of the predictions made.

\section{How explosions can affect binaries}
\label{sec:expl_bin}

While the majority of massive stars are born with companion(s), only a small
minority remain bound after the first core-collapse in the system \citep[e.g.][]{dedonder:97,
  eldridge:11, renzo:19walk}. For compact objects (i.e., neutron stars and black holes), being in a
binary system is the exception rather than the rule. This is counterintuitive, since binary
interactions are often the main or only way to observe the compact objects (e.g., through X-rays and/or
gravitational waves), especially in the case of black holes. In other words, the majority of
(isolated) massive binaries evolves to form
a single compact object and a ``widowed'' companion star.

The main reason why $86^{+11}_{-22}\%$ of binaries\footnote{This fraction excludes binaries that
  result in a stellar merger before the first CCSN.} are disrupted at the first core-collapse event
appears to be supernova (SN) natal kicks \citep[][]{renzo:19walk}. The width of the range reported above is
dominated by the uncertainties in the parametrization of the natal kicks, which allows to use the
population of ``widowed stars'' to observationally constrain the explosion physics of their former
companions in a statistical sense. In particular, the high-mass tail of the mass function of
``widowed stars'' is sensitive to the average black hole kick \citep[][]{renzo:19walk}: black holes
are produced by on average more massive stars, which typically have more massive companion that can
become single ``widowed stars'' if the black holes receive significant kicks at formation.

When
assuming spherical symmetry of the collapse and explosion in the frame of the exploding star (i.e.,
no natal kick), only $\sim\,16\%$ of binary systems are disrupted. In this cases, the change in
gravitational potential due to the rapid loss of the SN ejecta from the binary (so-called ``Blaauw
kick'', \citealt{blaauw:61}) alone unbinds the binary (see also \citealt{boubert:17a}). However,
typically Roche-lobe overflow will strip the envelope of the donor star which explodes first,
limiting the amount of mass that can be ejected at explosion.
Only for wide, non-interacting binaries the ``Blaauw kick'' is sufficient to separate the companions.

If the core-collapse produces a successful explosion, the newly ``widowed star'' will be hit by the blast
wave \citep[e.g.,][]{moriya:15b,hirai:18}, which can alter its appearance for a few thermal timescales by
depositing energy in the star and removing some mass, although the latter effect is typically small
\citep[e.g.][]{liu:15, rimoldi:16}.

Because of the binary disruption, the ``widowed star'' acquires a peculiar space velocity
corresponding to first order to its pre-explosion orbital velocity. Occasionally, this can produce
fast moving runaway stars \citep[e.g.,][]{blaauw:61, hoogerwerf:01}, however, it is much more common
that this peculiar velocity is relatively slow ($\sim$\,10\,$\mathrm{km\ s^{-1}}$), making them
``walkaway stars'' \citep[][]{renzo:19walk}. This happens because during the first stable Roche lobe
overflow, long before the SN
explosion, binaries tend to overall widen. Moreover, the mass transfer leads to an inversion of the
mass ratio. Both effects decrease the orbital velocity of the secondary. The velocity distribution
of ``widowed stars'', if observed, would add a constraint on the orbital evolution of massive binaries: their peculiar velocity with respect to their parent population
relates to how close to the companion they were at the time of the explosion.

\section{How binaries can affect the explosions}
\label{sec:bin_expl}

Binarity can have consequences on the rate and timing of CCSN events
\citep[e.g.,][]{podsiadlowski:92b, dedonder:03,zapartas:17}. For example, it can allow stars born
below the minimum mass to give a CCSN to ultimately explode (either because they accreted mass from
a companion, or because of mergers). \cite{zapartas:17} showed that $15^{+9}_{-8}\%$ of all CCSNe
might come from this type of evolutionary paths. These progenitor systems are generally longer-lived than
normal massive stars, resulting in delayed SNe compared to the age of a given (co-eval) parent
population. More convoluted evolutionary paths involving multiple phases of mass transfer (either stable or
unstable) might also possibly generate non-standard pre-explosion stellar structures resulting in peculiar
transients \citep[e.g.,][]{justham:14,menon:17}.

\subsection{Stripped-envelope SNe}

Binarity also impacts the observable properties of the stellar explosions themselves: as mentioned
above, the most common binary evolution path involves a phase of stable mass transfer which
typically removes the entire hydrogen-rich envelope of the donor star \citep[e.g.,][although this is
known to be Z-dependent]{kippenhahn:67, yoon:17,gotberg:17}. Therefore, typically the
first SN in the system will be a hydrogen-less type Ib or Ic SN, or possibly a IIb with only a
little amount of hydrogen remaining, commonly referred to all together as stripped-envelope SNe
\citep[e.g.,][]{smith:11, eldridge:13}.

\cite{zapartas:17b} showed that about $2/3$ of all stripped-envelope SNe are expected to occur in
the presence of a main sequence companion, assuming an initial mixture of single and binary stars
and sub-solar Z. This fraction drops below 1/2 only for parameter variations enhancing the wind mass
loss, including super-solar Z. These SNe are those unbinding the binary and creating the ``widowed'' stars.
Albeit common, the main sequence
companions might be challenging to find at the SN site because of their possibly low mass
\cite[][]{zapartas:17b}.

Most of the stripped-envelope SNe progenitors single at explosion were massive enough to get stripped
through winds, either because they were initially massive enough or because they accreted mass or
merged with a binary companion. In
$5^{+12}_{-4}\%$ of the cases, the exploding star had a compact object companion.

\subsection{H-rich SNe}

Perhaps more surprisingly, binary products might also contribute to a significant fraction of
hydrogen-rich SNe, despite the fact these can in principle be explained by single star evolution.
Single star models struggle to explain the variety of light curve morphology and spectral evolution
of these hydrogen-rich explosions. Using observationally motivated initial distributions,
\cite{zapartas:19} showed that accretion, and more importantly mergers in binaries can lead to stars
exploding with a significant amount of hydrogen left in their envelope. These SNe with a
binary-product progenitor could contribute to $45_{-12}^{+8}\%$ of all hydrogen-rich SNe (see also
\citealt{eldridge:19}). The ``widowed'' stars alone can contribute to $14_{-11}^{+4}\%$ of
hydrogen-rich SNe.

\section{Conclusions}

Massive binary evolution can proceed through a complex variety of paths depending on both initial
conditions and physical assumptions of the models. While theoretical understanding is far from final, the
exploration of the vast parameter space is becoming possible. Existing and upcoming observational constraints
from large surveys and gravitational waves are already guiding it. Loosely speaking, neglecting
stellar mergers, each massive binary will produce a close-to hydrogen-less donor star and a
hydrogen-rich accreting star. The outcome of stellar mergers depends on the evolutionary phases of
the two stars when they happen. Accounting for binarity when dealing with
observed samples of massive stars (even if presently single) and samples of CCSNe is important to
not misinterpret the observations.
\newpage
\acknowledgements
% Do not leave a blank line here! <---------------------->
We are really grateful for guidance and mentoring to S.~E.\ de~Mink, and to R.~G.~Izzard for
developing and maintaining the \texttt{binary\_c} code, and giving us access and support. We have
benefitted from uncountable enlightening discussions with F.~Broekgaarden, R.~Farmer, Y.~G\"otberg,
D.~Hendricks, S.~Justham, E.~Laplace, N.~Smith, K.~Temmink, S.~Toonen, A.~van~Son, D.~Vartanyan. EZ acknowledges support
from the Federal Commission for Scholarships for Foreign Students for the Swiss Government
Excellence Scholarship (ESKAS No. 2019.0091) for the academic year 2019-2020.

% \bibliography{./telc.bib}

\end{document}